\pdfoutput=1
\documentclass[aps,prb,letterpaper,reprint]{revtex4-1}

\usepackage{amsmath,amssymb}
\usepackage{bm}
\usepackage{textcomp}
\usepackage{lmodern}
\usepackage[pdftex]{graphicx}
\usepackage{textcomp}

\begin{document}

\title{Absence of the Pauli-Paramagnetic Limit in a Superconducting
U$_{\textrm{6}}$Co}

\author{Masahiro Manago}
\email{manago@scphys.kyoto-u.ac.jp}
\author{Kenji Ishida}
\affiliation{Department of Physics, Graduate School of Science,
Kyoto University, Kyoto 606-8502, Japan}

\author{Dai Aoki}
\affiliation{IMR, Tohoku University, Oarai, Ibaraki 311-1313, Japan}
\affiliation{INAC/PHELIQS, CEA-Grenoble, 38054 Grenoble, France}

\date{\today}
\begin{abstract}
We performed $^{59}$Co nuclear magnetic resonance (NMR) measurements of
single-crystalline U$_{\textrm{6}}$Co.
There is a small decrease in the Knight shift in the superconducting (SC)
state, but this change mainly arises from the SC diamagnetic effect.
The negligible change of the spin part of the Knight shift,
together with the absence of the Pauli-paramagnetic effect in the
SC U$_6$Co, is understood as a consequence  of the small spin susceptibility.
The nuclear spin-lattice relaxation rate $1/T_1$ is also measured in the
SC state under the magnetic field, and exhibits a tiny
Hebel-Slichter peak just below the SC transition temperature and
exponential behavior at lower temperatures.
These behaviors are in agreement with the full-gap $s$-wave pairing in U$_6$Co.
\end{abstract}

\maketitle

Uranium compounds U$_6$X (X = Mn, Fe, Co, and Ni) are superconductors with
relatively high superconducting (SC) transition temperatures $T_\text{c}$
among the uranium-based compounds.\cite{JPhysChemSol.7.259,PhysRev.168.464}
Contrary to other uranium-based heavy-fermion systems
including ferromagnetic superconductors,\cite{JPSJ.81.011003}
$5f$ electrons in U$_6$X compounds exhibit an itinerant nature
even at room temperature, and do not exhibit magnetic ground states.
The SC properties are consistent with the conventional one:
for instance,
the full-gap superconductivity of U$_6$Co ($T_\text{c} = 2.33$\,K) has been
indicated by the penetration depth,\cite{SolidStateCommun.61.101}
nuclear spin-lattice relaxation rate $1/T_1$,\cite{JPSJ.56.1645}
and more recently, by specific heat measurements.\cite{JPSJ.85.073713}
Thus, at present, these compounds are good reference systems for
the uranium-based unconventional superconductors.

It has been suggested that the superconductivity of U$_6$X might be close
to ferromagnetism,\cite{PhysRevB.31.7059}
because the total magnetic susceptibility of these compounds is so large
that it is comparable to that of the nearly ferromagnetic metal Pd.
This can be interpreted as a sign of the strongly enhanced electronic
spin susceptibility owing to the electronic correlation.
However, the magnetic properties of U$_6$X in the normal state are quite
different from those of ferromagnetic superconductors,\cite{JPSJ.81.011003}
in which ferromagnetic instability is commonly observed,
while it is usually unfavorable for conventional superconductivity.
Therefore, it is necessary to determine whether U$_6$X compounds
are close to the ferromagnetism and to consider the origin
of the large magnetic susceptibility in them.

It is also noteworthy that SC U$_6$Co has a large upper critical field
$H_\text{c2}$ for $T_\text{c}$.
In spin-singlet superconductors, the superconductivity is limited
by the Pauli-paramagnetic effect under a magnetic field.
This is because the spin susceptibility decreases in the SC state,
and the energy difference between the SC and the normal states reduces
under a larger magnetic field due to the magnetic energy in the normal state.
The Pauli-limiting field $H_\text{P}$ is expressed as
$\mu_0 H_\text{P}/\text{T} = 1.86 (2/g)T_\text{c}/\text{K}$
in the weak-coupling BCS model, where $g$ is the $g$-factor.
If this is applied to U$_6$Co assuming $g = 2$,
$\mu_0 H_\text{P} = 4.3$\,T is obtained.
However, $\mu_0 H_\text{c2}$ along the $[001]$ and $[110]$ directions in
U$_6$Co in the tetragonal structure is 7.85 and 6.56\,T,
respectively,\cite{JPSJ.85.073713} and both of them exceed 4.3\,T.
The Pauli-paramagnetic effect is actually absent in U$_6$Co.
The large $H_\text{c2}$ value has been considered to originate from the small
$g$-factor.\cite{JPSJ.85.073713}

In this paper, we report $^{59}$Co NMR measurements of U$_6$Co, which are
performed to further investigate the electronic spin susceptibility
from a microscopic point of view.
We found that the NMR Knight shift exhibits a small decrease in the SC state.
This result suggests that the spin susceptibility of U$_6$Co is not a dominant
term in the total susceptibility, but that the Van Vleck term is a main term.

We used a single-crystalline U$_{\textrm{6}}$Co sample with $T_\text{c} =
2.34$\,K for the NMR study.
The sample was cooled using liquid $^{4}$He or a $^{3}$He-$^{4}$He
dilution refrigerator.
$^{59}$Co NMR was performed with the magnetic field perpendicular to
the $[001]$ direction, and the Knight shift and $1/T_1$ were measured.
Since the $^{59}$Co nucleus has $I=7/2$ spin and the Co site does not have
cubic symmetry, the NMR spectrum splits into 7 lines
owing to a nuclear quadrupole interaction.
All the NMR data were measured at the central line
($1/2 \leftrightarrow -1/2$).
The magnetic field was calibrated using a $^{63}$Cu signal arising from
the NMR coil.
In addition to NMR measurements, we performed nuclear quadrupole resonance
(NQR) measurements of a powdered single-crystalline sample to obtain
information about the homogeneity of the crystal structure.

\begin{figure}
	\centering
	\includegraphics{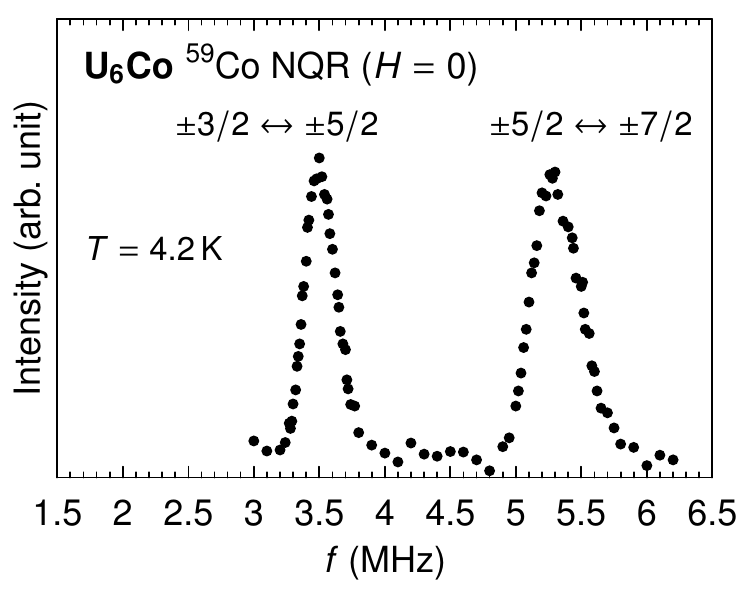}
	\caption{\label{fig:nqr}
	$^{59}$Co NQR spectrum for a powdered single-crystalline
	U$_6$Co sample under zero magnetic
	field in the normal state (4.2\,K).
	The left (right) peak arises from the transition of
	$\pm 3/2 \leftrightarrow \pm 5/2$ ($\pm 5/2 \leftrightarrow \pm 7/2$).
	}
\end{figure}

Figure \ref{fig:nqr} shows the $^{59}$Co NQR spectrum of the powdered sample.
The quadrupole frequency is estimated as $\nu_\text{Q} \simeq 1.77$\,MHz
for the present sample.
The asymmetric parameter $\eta$ is almost $0$, which is consistent with
the fact that the electric field gradient at the Co site has axial symmetry
with respect to the $[001]$ direction.
The value of $\nu_\text{Q}$ in this sample is slightly larger and the line
width is broader than those in the previous NQR study.\cite{JPSJ.56.1645}
Nevertheless, a clear SC anomaly is observed in the single-crystalline sample,
as mentioned later, and thus, the present samples are suitable for studying the
superconductivity of U$_6$Co.

\begin{figure}
	\centering
	\includegraphics{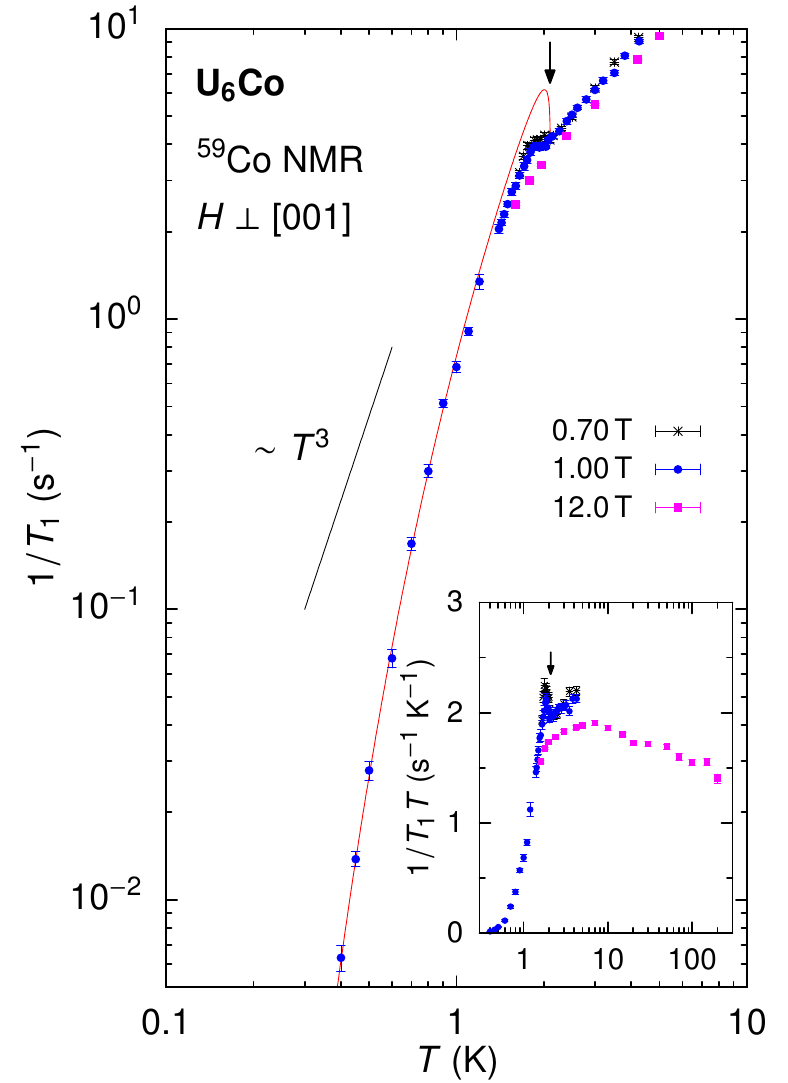}
	\caption{\label{fig:t1}
	(Color online) $^{59}$Co nuclear spin-lattice relaxation rate $1/T_1$
	of a single-crystalline U$_6$Co sample at 0.70, 1.00 and 12.0\,T
	perpendicular to the $[001]$ direction.
	The third magnetic field exceeds $H_\text{c2}$.
	The solid line is the calculated curve of SC $1/T_1$ with
	$2\Delta(0) / k_\text{B}T_\text{c} = 3.7$ and $\delta/\Delta(0) = 0.44$.
	$T^3$ dependence is expected in the case of the line-node gap,
	and the actual $1/T_1$ decreases faster than that shown.
	The inset shows the temperature dependence of $1/T_1 T$.
	The field dependence of $1/T_1 T$ could be ascribed to
	a tiny amount of magnetic impurity and/or
	the structure of the density of states around the Fermi energy.
	The arrows indicate $T_\text{c}$ at 1.00\,T (2.10\,K).}
\end{figure}

\begin{figure}
	\centering
	\includegraphics{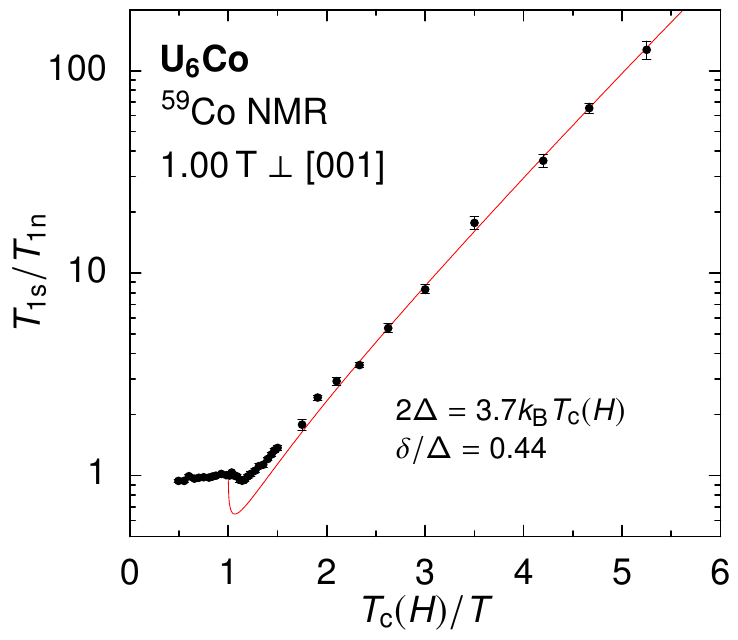}
	\caption{\label{fig:t1-2}
	(Color online) Plot of $T_{1\text{s}}/T_{1\text{n}}$
	against $T_\text{c}(H)/T$ at 1.00\,T.
	The linear relation in this plot indicates the exponential decrease of
	$1/T_1$.
	The solid line is identical to that in Fig.~\ref{fig:t1}.}
\end{figure}

The $1/T_1$ values in U$_{\textrm{6}}$Co at 1.00\,T are shown in
Fig.~\ref{fig:t1}.
A Hebel-Slichter (HS) peak is found just below $T_\text{c}$
as observed in the previous NQR measurement under zero magnetic
field,\cite{JPSJ.56.1645}
although the HS peak is more strongly suppressed than that in the previous NQR.
This result again indicates that U$_{\textrm{6}}$Co is an $s$-wave
superconductor.
The HS peak is slightly larger at 0.70\,T than at 1.00\,T, indicating that
the peak is suppressed by the magnetic field.
Figure \ref{fig:t1-2} shows $T_{1\text{s}}/T_{1\text{n}}$ against
$T_\text{c}(H)/T$, where $1/T_{1\text{s}}$ ($1/T_{1\text{n}}$) is the
relaxation rate in the SC (normal) state.
An exponential decrease of $1/T_1$ is observed at lower temperatures,
although the relaxation curves have a short-$T_1$ component in the SC state
arising from the vortex core.
The $1/T_1$ value shown in Figs.~\ref{fig:t1} and \ref{fig:t1-2} corresponds
to a longer-$T_1$ component.
The $1/T_1$ behavior in U$_6$Co is evidence of a nodeless gap.

The solid curves in Figs.~\ref{fig:t1} and \ref{fig:t1-2} are a result of
the calculation based on the $s$-wave full-gap model, and they reproduce
the experimental result in $T \lesssim 0.5T_\text{c}$.
The density of states (DOS) in the SC state is broadened by a
rectangle function with a width of $2\delta$ and a height of $1/2\delta$
for the calculation of $1/T_1$.
The calculated $1/T_1$ and DOS are expressed as follows:
\begin{align*}
	\frac{T_{1\text{n}}}{T_{1\text{s}}} &= \frac{2}{k_\text{B}T}
	\int _0 ^{\infty} [N_\text{s}(E)]^2 \left( 1 + \frac{\Delta^2}{E^2} \right)
	f(E) [ 1 - f(E)] \, dE,\\
	N_\text{s}(E) &= \frac{1}{2\delta} \int _{E-\delta} ^{E+\delta}
	\frac{E'}{(E'^2 - \Delta^2)^{1/2}} \, dE'.
\end{align*}
The SC gap $\Delta$ is set to $2\Delta / k_\text{B} T_\text{c}(H)
= 3.7$ from Ref.~\onlinecite{JPSJ.85.073713}, and $\delta/\Delta = 0.44$.
The strong suppression of the HS peak just below $T_\text{c}$
is not reproduced even with the relatively large $\delta$ and may be ascribed
to the magnetic field.
Recently, the Doppler shift effect, or the so-called Volovik effect, has been
proposed as a mechanism of the suppression of the HS peak
of the $s$-wave superconductor under the magnetic
field.\cite{PhysRevB.93.140502}
The Volovik effect has been regarded as a minor effect on the low-energy
excitation of quasiparticles in the case of full-gap superconductors,
but this cannot be neglected for the HS peak, because it is
sensitive to the edge form of the SC DOS particularly just below $T_\text{c}$.
Therefore, it is considered that the suppression of the peak in U$_6$Co
under the magnetic field could be explained by the Volovik effect.

$1/T_1$ of U$_6$Co in the normal state exhibits the Korringa relation with
$1/T_1T \simeq 2.0\, \text{s}^{-1}\text{K}^{-1}$ just above $T_\text{c}$,
as seen in the inset of Fig.~\ref{fig:t1}, and this value is not strongly
enhanced compared with other uranium-based heavy-fermion systems.
For instance, this is quite smaller than that of the itinerant
ferromagnetic superconductor UCoGe: $1/T_1T \sim 10^{2}$--$10^3\,
\text{s}^{-1}\text{K}^{-1}$ in the paramagnetic state under zero magnetic
field depending on the temperature.\cite{JPSJ.79.023707}
Instead, $1/T_1T$ of U$_6$Co is close to that of $^{59}$Co $1/T_1T$ in YCoGe,
which is a nonmagnetic reference compound of UCoGe:
$1/T_1T = 2.03\, \text{s}^{-1}\text{K}^{-1}$ (Ref.~\onlinecite{JPSJ.80.064711}).
The ordinary $1/T_1T$ value in the normal state implies that the
SC phase of U$_6$Co is not in proximity to the ferromagnetic ground state.

\begin{figure}
	\centering
	\includegraphics{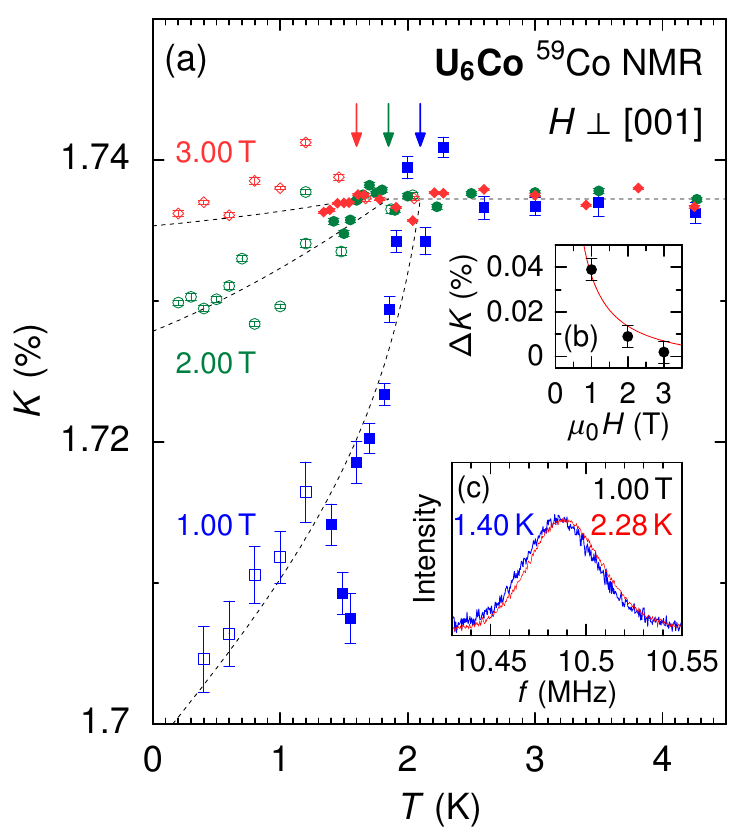}
	\caption{\label{fig:knightshift}
	(Color online) (a) $^{59}$Co-NMR Knight shift in U$_6$Co with the field
	perpendicular to the $[001]$ direction.
	The arrows indicate $T_\text{c}$ in each field.
	The open (closed) symbol represents the result with a
	dilution refrigerator (standard $^{4}$He pumping).
	The Knight shifts shown here are the high-field limit values
	for eliminating the apparent field dependence
	even after subtracting the quadrupole effect.
	The apparent Knight shift $K_\text{obs}$ in the normal
	state measured at the central line ($1/2 \leftrightarrow -1/2$)
	depends on the field $H$ as $K_\text{obs} = K + c/H^2$, where $c$ is
	a constant, and this extrinsic variation is caused by a small error
	of the nuclear quadrupole frequency $\nu_\text{Q}$ and the nonzero value
	of the square of the asymmetric parameter of the electric field gradient
	$\eta^2$.
	The dashed lines are the guides to eye.
	The scattering of the data points mainly originates from the time-dependent
	field drift of the SC magnet, which is estimated as $\sim 0.1$\,mT.
	(b) The Knight-shift change in the SC state
	$\Delta K \equiv K(T>T_\text{c}) - K(0)$ against $H$.
	The solid line is a theoretical curve\cite{PhysRevB.68.054506}
	of the diamagnetic shift $K_\text{dia}$ with $\kappa = 86$ (see text).
	(c) The $^{59}$Co-NMR spectra at 1.00\,T at the central line.
	The left (right) spectrum is measured at 1.40\,K (2.28\,K)
	in the SC (normal) state.}
\end{figure}

The Knight shifts of U$_{\textrm{6}}$Co at several fields below 4.2\,K
are shown in Fig.~\ref{fig:knightshift} (a), and the NMR spectra at 1.00\,T
are shown in Fig.~\ref{fig:knightshift} (c).
The decrease in the Knight shift extrapolated to $T = 0$
is $\simeq 0.039 \pm 0.005$\% (1.00\,T), $0.009 \pm 0.005$\% (2.00\,T),
and $0.002 \pm 0.005$\% (3.00\,T),
and these values are much smaller than the total Knight shift of
${\sim} 1.7$\%.
In general, the Knight shift decreases due to a diamagnetic field
$H_\text{dia}$ as well as a decrease in the spin susceptibility in the SC state.
$H_\text{dia}$ is roughly expressed as $H_\text{dia} \simeq H_\text{c1}
\log(H_\text{c2}/H) /(2 \log \kappa)$ (defined as positive),
where $H_\text{c1} \ll H \ll H_\text{c2}$,
$H$ is the external field, $\kappa$ is a Ginzburg-Landau parameter,
and $H_\text{c1}$ is the lower critical field.\cite{Tinkham}
Therefore, the diamagnetic shift $K_\text{dia} = H_\text{dia}/H$ is strongly
suppressed with increasing $H$.
On the other hand, the spin part of the Knight shift is
$K_\text{s}(0)/K_\text{s}(T>T_\text{c}) \simeq a H/H_\text{c2}$ at lower fields
($a$ is close to unity), and thus the Knight-shift change arising from
the spin part does not strongly depend on $H$ as long as $H/H_\text{c2} \ll 1$.
The experimental Knight-shift decrease is suppressed with increasing $H$
even well below $H_\text{c2}$ and this is qualitatively similar to the
behavior of $K_\text{dia}$.
In addition, the field dependence of the Knight-shift decrease is roughly
reproduced with a theoretical formula of the diamagnetic field,
which is valid in a wide range of the fields,\cite{PhysRevB.68.054506}
as shown in Fig.~\ref{fig:knightshift} (b).
Here, we adopt $\kappa = 86$ for the best fit of the experiment, which
is close to $\kappa \simeq 70$ estimated from $H_\text{c2}$ and the
thermodynamic field.\cite{JPSJ.85.073713}
This also suggests that the diamagnetic effect is a dominant origin of
the Knight-shift change in U$_{\textrm{6}}$Co.

The magnitude of $K_\text{s}$ could not be properly determined in this study,
but there is a constraint on the value of $K_\text{s}$, as discussed below.
Noting that $K_\text{dia}$ decreases faster than $H^{-1}$ against $H$,
and assuming that $a \simeq 1$ for
$K_\text{s}(0)$, then $K_\text{s}$ in the normal state satisfies
$-0.05\% \lesssim K_\text{s} \lesssim 0.01\%$.
The lower limit is not quite strict because the concrete
form of $H_\text{dia}(H)$ is not assumed, and the upper limit is
obtained by the Knight-shift change at 3.00\,T.
Therefore, this constraint on $K_\text{s}$ does not necessarily mean that
a negative $K_\text{s}$ is more likely than a positive one.

Here, we attempt to estimate how the Knight shift would behave
when the Pauli-paramagnetic effect is absent in U$_6$Co.
Since the magnetic susceptibility is temperature-independent in the normal
state,\cite{JPSJ.85.073713}
the hyperfine coupling constant $A_\text{hf}$ cannot be determined
through a comparison between the susceptibility and the Knight shift.
Thus, we tentatively adopt $A_\text{hf} \simeq 3\,\text{T}/\mu_\text{B}$
for the $^{59}$Co hyperfine coupling constant in U$_6$Co, which is a
typical value for $^{59}$Co in uranium-based compounds such as
UCoGe\cite{PhysRevLett.105.206403} and UCoAl.\cite{JPSJ.80.093707}
This is based on the assumption that the U $5f$ electrons are transferred
to the Co $4s$ electrons, and the contribution of the Co $3d$ electrons is
relatively small.
Considering the absence of the Pauli-paramagnetic effect,
$H_\text{P}$ satisfies $H_\text{P} = H_\text{c} / \sqrt{\Delta\chi_\text{s}}
> H_\text{c2}$,
where $\Delta \chi_\text{s}$ is the decrease of the spin susceptibility
in the SC state (a volume susceptibility in SI units for calculating
$H_\text{P}$) and $H_\text{c}$ is the thermodynamic field.
Using $\mu_0 H_\text{c} = 0.065$\,T at 0\,K (Ref.~\onlinecite{JPSJ.85.073713})
and the lattice parameters\cite{JPhysChemSolids.36.123} of U$_6$Co,
$\Delta \chi_\text{s} \lesssim 0.7 \times 10^{-4}\,\text{emu}/\text{U-mol}$
is deduced, which is much smaller then the total value
$\chi = 5.3 \times 10^{-4}\,\text{emu}/\text{U-mol}$.\cite{JPSJ.85.073713}
Thus the Knight-shift change in a spin part is estimated as
$\Delta K_\text{s} = A_\text{hf} \Delta \chi_{\text{s}} / N_\text{A}
\mu_\text{B} \lesssim 0.04$\%, where $N_\text{A}$ is Avogadro's number,
and $\mu_\text{B}$ is the Bohr magneton.
The actual change of the Knight shift after subtracting the diamagnetic
effect is sufficiently less than this limit.

It is known that the Knight-shift decrease is suppressed by spin-orbit
scattering in the dirty-limit superconductors.%
\cite{PhysRevLett.3.325,PhysRevB.62.11363}
In the case of U$_{\textrm{6}}$Co, since the mean free path is estimated to
be on the same order of the BCS coherence
length,\cite{JPSJ.85.073713} this mechanism plays a minor role in the
suppression of the Knight shift change.
Although the above discussion is based on the assumption that $A_\text{hf}$ is
positive and not very small, the small spin susceptibility is a promising
origin of the small Knight-shift change in U$_{\textrm{6}}$Co.
The absence of the Pauli-paramagnetic effect is closely related to
the Knight-shift behavior in the SC state in U$_6$Co.

In some uranium-based superconductors
such as UPt$_3$\cite{PhysRevLett.80.3129}
and URu$_2$Si$_2$\cite{JPSJ.85.073711}, the Knight-shift change due
to SC pairing is much smaller than the total value.
This means that the large part of the Knight shift is not related to
the quasiparticles at the Fermi energy $E_\text{F}$.
This also applies to U$_6$Co: $1/T_1T$ slightly depends on the temperature
as seen in the inset of Fig.~\ref{fig:t1}, reflecting the structure of
the DOS around $E_\text{F}$,
while such a temperature-variation is not seen at all in the magnetic
susceptibility of U$_6$Co\cite{JPSJ.85.073713} and in
the present Knight shift at 12.0\,T (not shown).
This indicates that the spin part of the Knight shift and spin susceptibility
are quite small in total.
Therefore, the spin state of the SC pairing should be carefully discussed when
the Knight-shift change is absent in uranium-based superconductors.

\begin{figure}
	\centering
	\includegraphics{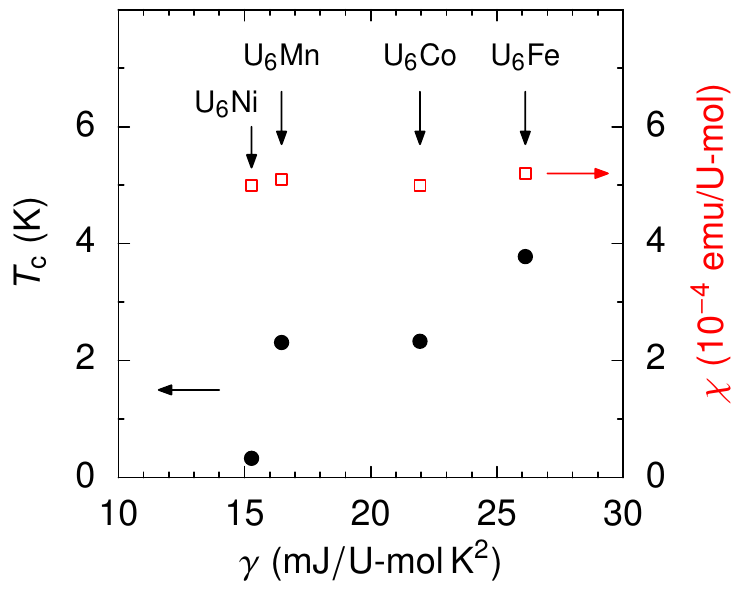}
	\caption{\label{fig:U6X}(Color online)
	Plot of SC transition temperatures $T_\text{c}$ (left, closed circle)
	and magnetic susceptibilities $\chi$ (right, open square) against
	electronic specific heat coefficients $\gamma$ in U$_6$X
	compounds.\cite{PhysRevB.31.7059}
	}
\end{figure}

The small spin susceptibility of U$_6$Co is also inferred from a
comparison among U$_6$X compounds.
The discussion here is based on Table VII in Ref.~\onlinecite{PhysRevB.31.7059},
and some of the data are shown in Fig.~\ref{fig:U6X}.
The electronic specific heat coefficient of U$_6$X varies between
$\gamma = 15$--$26\, \text{mJ}/(\text{K}^2\, \text{U-mol})$,
and the compound with a larger $\gamma$ has a higher $T_\text{c}$.
This tendency is consistent with the BCS theory, in which a larger
DOS at $E_\text{F}$ is more favorable for the superconductivity.
However, the total susceptibilities of U$_6$X compounds are
similar (approximately $5 \times 10^{-4}\,\text{emu}/\text{U-mol}$
in the powder samples), and this suggests that most of the susceptibility
is independent of the quasiparticles around $E_\text{F}$.
It seems that the Van Vleck term should make a dominant contribution to
the susceptibility in these systems.
The small spin susceptibility is
consistent with the small $g$-factor,\cite{JPSJ.85.073713}
and this is not surprising if the strong spin-orbit
coupling in actinides is taken into consideration.\cite{RevModPhys.81.235}

Finally, we comment on the present Knight shift in U$_6$Co.
A positive $A_\text{hf}$ and thus, positive $K_\text{s}$ are
assumed in the above discussion for simplicity,
but this is not actually evident since the Co $3d$-electronic spins would lead
to negative hyperfine coupling owing to core polarization if they have finite
DOS at $E_\text{F}$.
If the core polarization is a dominant mechanism in the spin part
of the Knight shift, $A_\text{hf}$ should be negative.
In this case, the Knight shift could shift to larger values in the SC state
depending on the magnitude of $A_\text{hf}$, which is not experimentally
detected in the present field ranges.
There is even a possibility that the positive Co $4s$ and negative $3d$
contributions cancel out each other,
as seen in V metal,\cite{RevModPhys.36.177}
and the spin part of the Knight shift might
remain constant in the SC state even if the spin susceptibility
is sufficiently large for the Pauli-paramagnetic effect to occur.
As for the orbital part of the $^{59}$Co Knight shift in U$_6$Co, this does
not necessarily originate only from the U $5f$ Van Vleck terms.
Co $3d$ electrons also produce the orbital part of the Knight shift
depending on the band occupancy.
These issues can be resolved if the band structure of U$_6$Co is clarified.

In summary, we performed $^{59}$Co NMR measurements on a single
crystal of U$_{\textrm{6}}$Co, and found that the Knight-shift change in
the SC state is small compared with the total Knight shift.
This result, as well as the large upper critical field, can be understood
to occur because the spin susceptibility is much smaller than the
total susceptibility.
The absence of the Pauli-paramagnetic effect is likely to be associated with
the negligible change in the Knight shift in this system.
The value of the expected Knight-shift change in the SC state was estimated by
using the plausible value of the hyperfine coupling constant.
The $1/T_1$ behavior in the SC state is in good agreement with
the $s$-wave full gap of U$_6$Co.

The authors would like to thank K. Okamoto and T. Urai for contributing to
the experiment.
They would also like to thank Y. Yanase, S. Kitagawa, Y. Maeno,
S. Yonezawa, H. Harima, and H. Ikeda for their valuable discussions.
One of the authors (MM) is a research fellow of the
Japan Society for the Promotion of Science (JSPS).
This work was supported by Kyoto University LTM Center, and by
a Grant-in-Aid for Scientific Research (Grant No.~JP15H05745) and
Grant-in-Aids for Scientific Research on Innovative Areas ``J-Physics''
(Grants No.~JP15H05882, No.~JP15H05884, and No.~JP15K21732) from JSPS.

\end{document}